\documentclass[letter]{aa}
\usepackage{graphicx}
\usepackage{txfonts}
\newcommand{\Msun}{\mbox{$\rm M_{\odot}$}}
\newcommand{\Msunyr}{\mbox{$\rm M_{\odot}$\,yr$^{-1}$}}

\newcommand{\Rsun}{\mbox{$\rm R_{\odot}$}}

\newcommand{\micron}{$\mu$m}

\def\mathstacksym#1#2#3#4#5{\def#1{\mathrel{\hbox to 0pt{\lower#5\hbox{#3}\hss} \raise #4\hbox{#2}}}}
\mathstacksym\gta{$>$}{$\sim$}{1.5pt}{3.5pt} 
\mathstacksym\lta{$<$}{$\sim$}{1.5pt}{3.5pt} 
\begin{document}
                                %
\title{Mid-infrared interferometry towards the massive young stellar object CRL\,2136: inside the dust rim\thanks{Based on observations with the VLTI, proposal 381.C-0607}}
 
\titlerunning{MIDI measures a disk near CRL2136}
\authorrunning{de Wit et al.}
\author{W.J. de Wit\inst{1,2}, M.G. Hoare\inst{2}, R.D. Oudmaijer\inst{2}, D.E.A N\"urnberger\inst{1}, H.E. Wheelwright\inst{2}, S.L. Lumsden\inst{2}} 
\offprints{W.J. de Wit, \email{wdewit@eso.org}}
\institute{European Southern Observatory, Alonso de Cordova 3107, Vitacura, Santiago, Chile \and
School of Physics \& Astronomy, University of Leeds, Woodhouse Lane, Leeds LS2 9JT, UK}

\date{Received date; accepted date}
\abstract
{Establishing the importance of circumstellar disks and their properties is crucial to fully understand massive star formation. }
{We aim to spatially resolve the various components that make-up the accretion environment of a massive young stellar object ($\la 100$\,AU), and 
 reproduce the emission from near-infrared  to millimeter wavelengths using radiative transfer codes.}
{We apply mid-infrared spectro-interferometry to the massive young stellar object CRL\,2136. The
observations were performed with the Very Large Telescope Interferometer and the MIDI instrument at a 42\,m baseline probing angular scales of 50 milli-arcseconds. We model the 
observed visibilities in parallel with diffraction-limited images at both 24.5\,$\mu$m and in the N-band (with resolutions of 0.6\arcsec and 0.3\arcsec, respectively),  
as well as the spectral energy distribution. } 
{The arcsec-scale spatial information reveals the well-resolved emission from the dusty envelope. By simultaneously 
 modelling the spatial and spectral data, we find that the bulk of the dust  emission occurs at several dust sublimation radii (approximately 170 AU). 
This reproduces the high mid-infrared fluxes and at the same time the low visibilities observed in the MIDI data for 
wavelengths longward of 8.5\,\micron. However, shortward of this wavelength the visibility data show a sharp up-turn indicative of 
compact emission. We discuss various potential sources of this emission. We exclude a dust disk being responsable for the observed spectral 
imprint on the visibilities. A cool supergiant star and an accretion disk are considered and both shown to be viable origins of the compact mid-infrared emission.
}
{We propose that CRL\,2136 is embedded in a dusty envelope, which truncates at several times the dust sublimation radius. 
A dust torus is manifest in the equatorial region. We find that the spectro-interferometric $N$-band signal can be reproduced 
by either a gaseous disk or a bloated central star. If the disk extends to the stellar surface, it accretes at a rate of $3.0\,10^{-3}$\,\Msunyr.}

\maketitle
\section{Introduction}
Mid-infrared stellar interferometry of young massive stars has the potential to
bring significant new insights into the process of massive star
formation. The technique overcomes the two age-old problems of 
low angular resolution and high extinction. In de Wit et
al. (2007\nocite{2007ApJ...671L.169D}, 2010\nocite{2010A&A...515A..45D}), we presented spectrally dispersed 
mid-infrared (mid-IR) interferometric observations of the massive young stellar object (MYSO) W33A 
using the Very Large Telescope Interferometer (VLTI).  We found that the $N$-band emission is dominated by warm dust on 100\,AU scales,
located in the walls of the outflow cavity. Recently, VLTI observations in the near-infrared (near-IR) 
using the AMBER instrument show unambiguously  the presence of a disk-like structure 
on size scales of $\sim 15$\,AU (Kraus et al. 2010\nocite{2010Natur.466..339K}).

In this letter, we focus on the MYSO CRL\,2136. The dominant 
infrared source IRS\,1 ($L\sim7\,10^{4}L_{\odot}$) is the driving force of an
arcminute scale bipolar CO outflow with a P.A.  of $\sim 135\degr$ (Kastner
et al. 1994\nocite{1994ApJ...425..695K}). Weak, optically thick radio emission
originating from IRS\,1  was detected by Menten \& Van der Tak (2004\nocite{2004A&A...414..289M}).  Near-IR polarimetric
observations demonstrate the presence of a polarization
disk (Minchin et al. 1991\nocite{1991MNRAS.251..522M}; Murakawa et al. 2008\nocite{2008A&A...490..673M}). 
These credentials render CRL\,2136 a promising target to probe for the geometry of the material on milli-arcsecond angular scales 
in the harsh environment close to the stellar surface. We perform VLTI observations in the
$N$-band, where MYSOs are bright and warm dust in the envelope and possibly in a circumstellar disk should still 
 contribute significantly to the total flux.

\section{Observations and data reduction}
\label{Observations}

\begin{table*}
  {
     \begin{center}
      \caption[]{VLTI observational details and basic source properties.}
      \label{tabobs}
      \begin{tabular}{ccccccccccc}
        \hline
        \hline
          Name  & RA & Dec & D & Lum & date   & Stations     & B      & P.A. & AM   & Int. Calibrator  \\
                     & (h:m:s)  & ($\degr$:$\arcmin$:$\arcsec$)       & (Kpc)  & ($\rm L_{\odot}$)&     &              & (m)    & (\degr) &&            \\
        \hline
        CRL\,2136 & 18:22:26.3 & $-$13:30:12.0 & 2.0& $7\,10^{4}$ & 23-06-08  & U2-U3        & 42.57 & 47.45  &  1.4& HD\,175775\\
        \hline
        \hline
      \end{tabular}
    \end{center}
  }
\end{table*}
\begin{figure*}
  \includegraphics[height=6cm,width=6cm,angle=90]{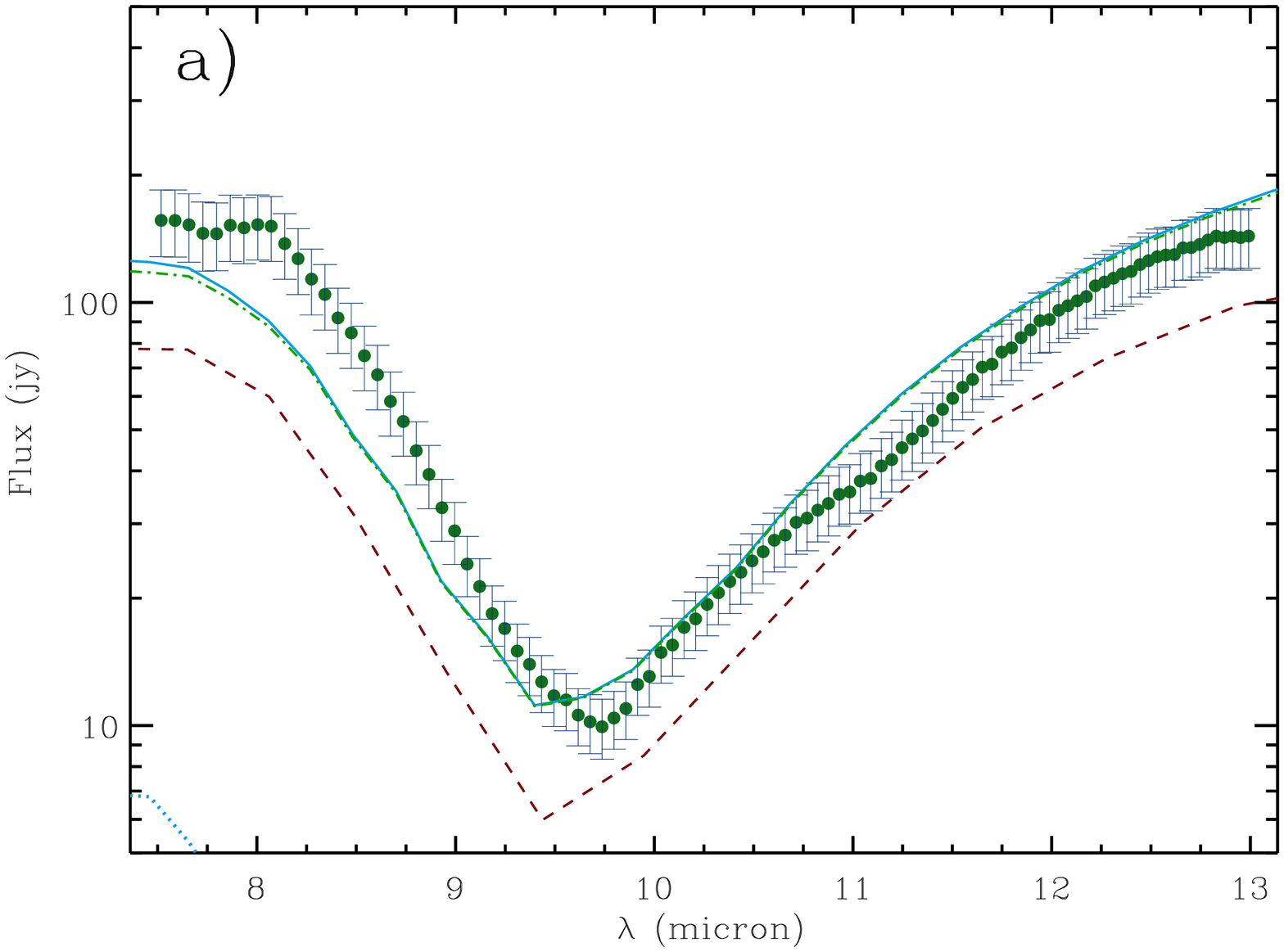}
  \includegraphics[height=6cm,width=6cm,angle=90]{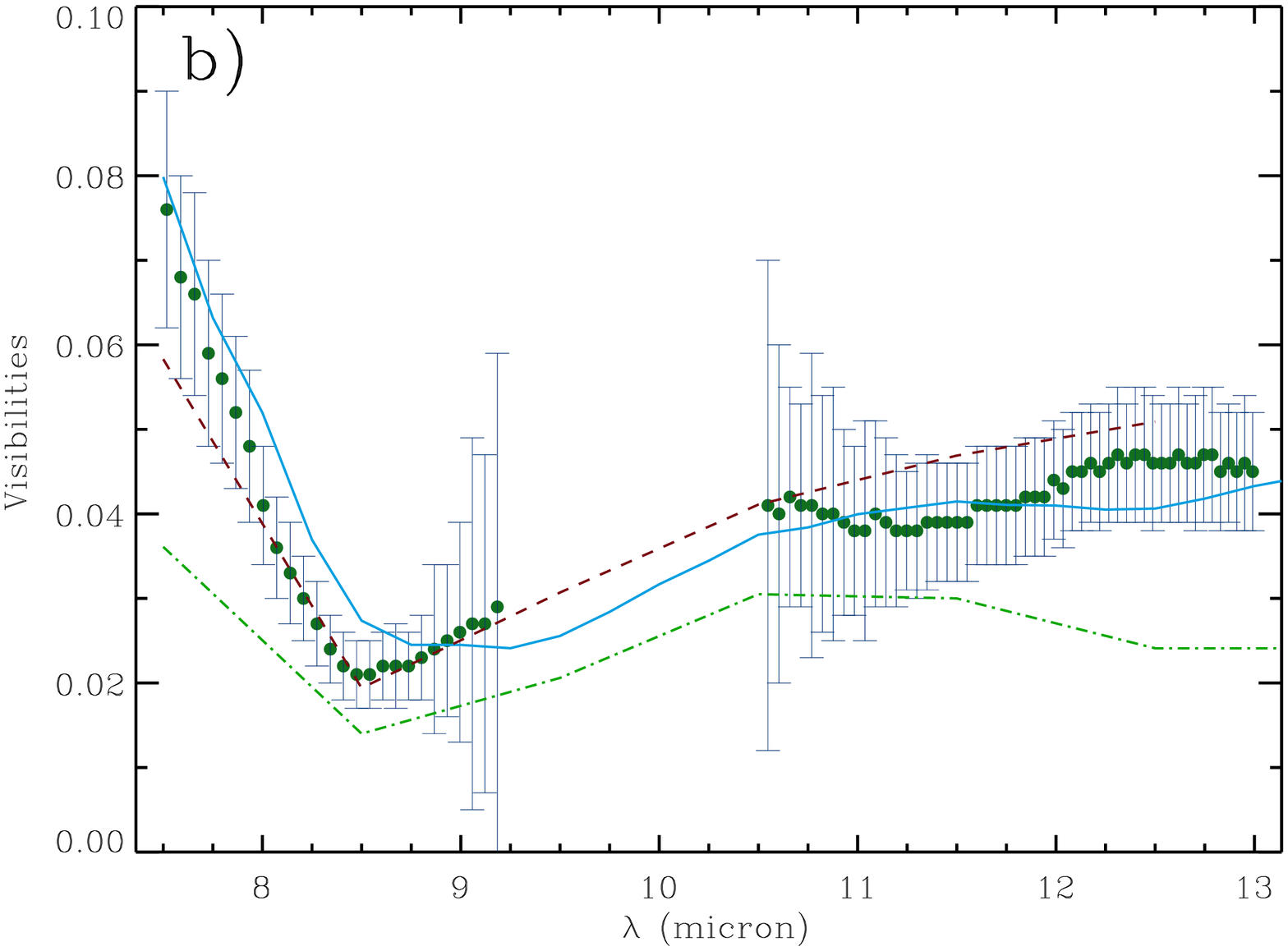}
 \includegraphics[height=6cm,width=6cm,angle=90]{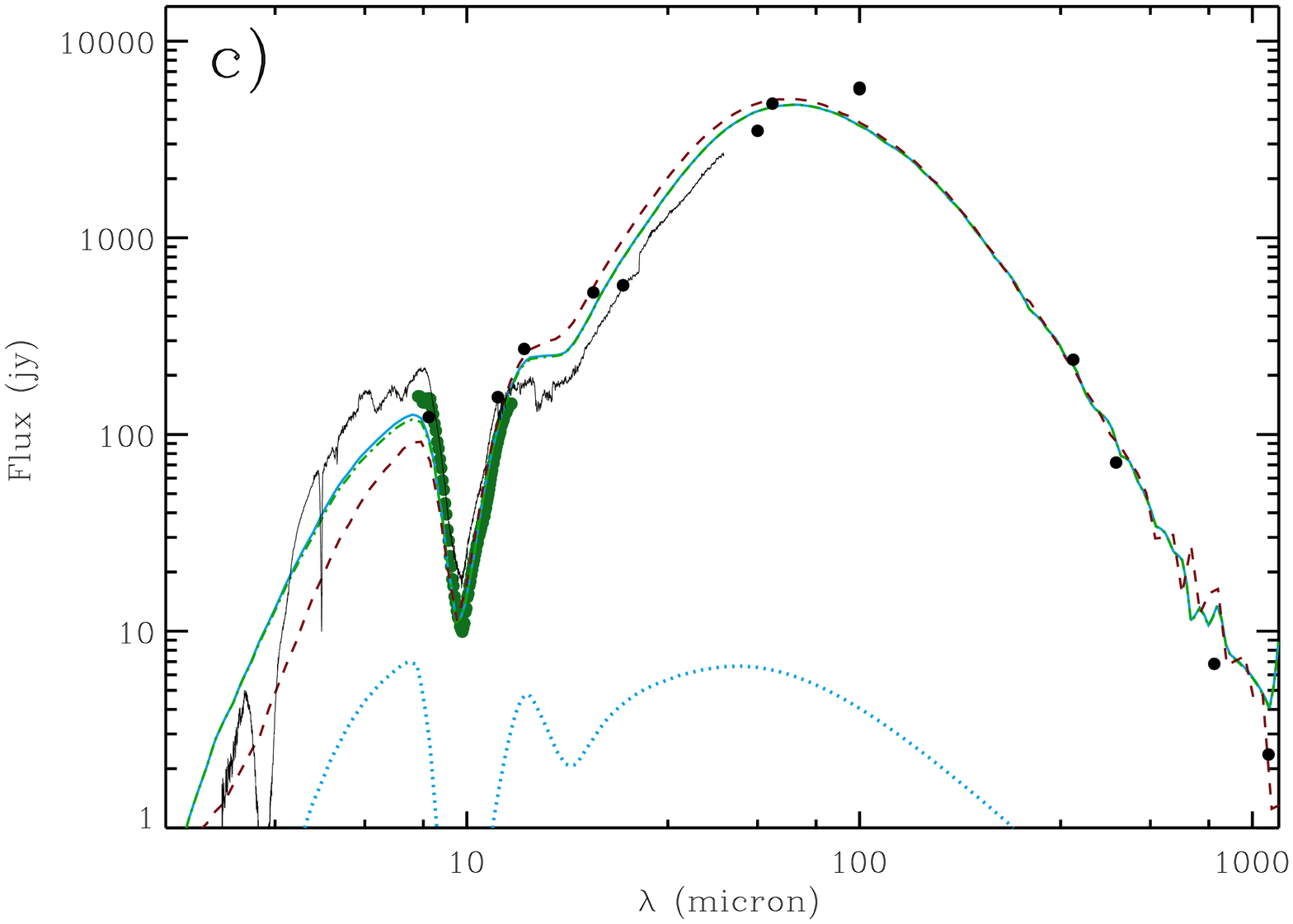}
  \caption[]{Observations (dots) and models (lines)  presented in
      Sect.3.2.  {\bf a) }$N$-band flux spectra; {\bf b)}
    Visibilities, observations affected by telluric absorption are
    omitted; {\bf c)} SEDs, with observations taken from de Wit et
    al. (2009\nocite{2009A&A...494..157D}) compilation. Graph styles
    are the same for each panel: 1) green dash-dotted line - ZAMS star
    $+$ envelope; 2) red dashed line - cool hypergiant $+$ envelope;
    3) blue full line - ZAMS star $+$ envelope and a gaseous
    disk. Panel c) shows the spectrum of the accretion disk
    (blue dotted line) and the observed ISO-SWS spectrum (full black line). Models that 
fit the SED only (Sect. 3.1) are not shown here.

}
  \label{spec}
\end{figure*}
The details of the target observations and VLTI 
configuration are presented in Table\,\ref{tabobs}. The VLTI baseline P.A. is perpendicular to the outflow direction ($\sim 135\degr$). The
angular resolution is about 50 milli-arcseconds or 100\,AU for the adopted distance of 2\,kpc. The observations were
executed with the VLTI mid-IR 2-beam combiner MIDI in its {\it High-Sens} MIDI mode. It uses all
the incoming light for the interferometric observation providing a measurement of the correlated flux.
A prism with a spectral resolution of 30 was employed to disperse the
incoming beams. A flux spectrum was taken immediately after the
interferometric observation. We estimate there to be a 15\% background variation between the observations as 
 indicated by the observed flux levels of interferometric calibrator stars monitored throughout the 
night. The change in the transfer function (i.e. instrumental visibilities) was also estimated using these stars, which led to a small correction in the final calibrated visibilities. 
The calibrator was also used as an 
approximate flux calibration (Cohen et al. 1999\nocite{1999AJ....117.1864C}). Detailed descriptions of the MIDI observational
procedure are given in Przygodda et
al. (2003\nocite{2003Ap&SS.286...85P}),  Leinert et
al. (2004\nocite{2004A&A...423..537L}), and Chesneau et
al. (2005\nocite{2005A&A...435..563C}). Data reduction was performed using the MIA$+$EWS software
package (version 1.6; see Jaffe 2004\nocite{2004SPIE.5491..715J};
K\"{o}hler 2005\nocite{2005AN....326Q.563K}), in which interferograms are
added coherently to maximize the signal-to-noise ratio. 

The measured MIDI fluxes and visibilities are presented in Fig.\,\ref{spec}. 
Panel\,1a shows the distinct silicate absorption of MYSOs. The flux error bars  are estimates of the fluctuations of the chopped frames and the
systematic 15\% sky uncertainty added in quadrature. Panel\,\ref{spec}b presents the calibrated visibilities. Correlated fluxes are 
compromised by telluric absorption at wavelengths shortwards of 7.5\,\micron, longwards of 13.3\,\micron, and between 9.3 and 10.2\,\micron, 
as determined from the calibrator spectrum. These wavelength ranges are therefore omitted in Fig.\,\ref{spec}b.
The error bars associated with the visibility points take into account the uncertainties in both the flux and the correlated flux; the latter uncertainty includes the error introduced by detector noise.
The visibilities are highest at the short wavelength end, then decrease to a minimum 
and gently rise again at longer wavelengths. A similar spectral behaviour is observed for the 
candidate MYSO IRS\,9A in the NGC\,3603 cluster (Vehoff et al. 2010\nocite{2010A&A...520A..78V}) and found to
be incompatible with a spherical dust distribution.  When
we interpret the visibilities as being caused by simple Gaussian-shaped emission regions, 
then CRL\,2136 quickly increases in size from 30\,mas (7.5\,\micron) to 42\,mas (8.5\,\micron) and then more gradually 
to 57\,mas (13\,\micron). Although there is a minimum in the  visibilities, the equivalent Gaussian sizes steadily 
increase because of the increasing wavelength. 

\section{Modelling and discussion}
\subsection{Excluded scenarios}
\label{discussion}
\begin{figure}
  \includegraphics[height=9cm,width=7cm,angle=90]{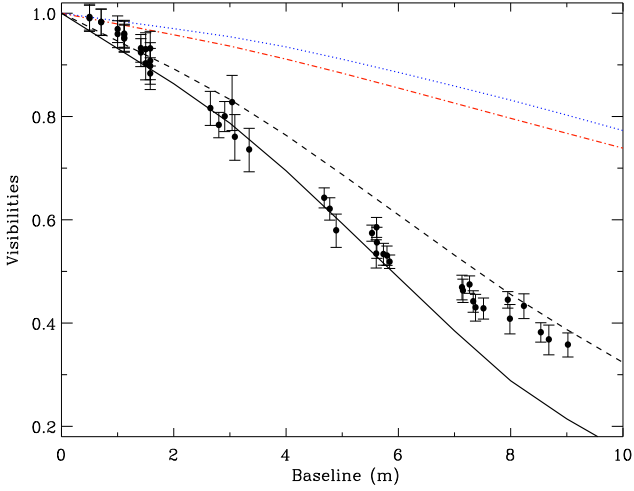}
  \caption[]{Short-spacing $N$-band visibilities against baseline for the envelope  model of Sect.\,3.2  at two perpendicular PAs (full line, and dashed line) 
compared to data from Monnier et al. (2009). We also indicate the predictions of two dust disk+envelope
models that fit the SED only of Sect. 3.1  (blue dotted and red dash-dotted).}
  \label{ste}
\end{figure}

The MIDI visibility spectrum provides evidence of a  steep drop in 
relative flux contribution with wavelength from a compact component.
As an approximation of the MYSO envelope emission (e.g. de Wit et
al. 2007\nocite{2007ApJ...671L.169D}), simple spherical models as are unable to  reproduce this visibility trend. 
However, the use of these models is justified by a broad-band 10\,\micron\,image produced by the 
Keck segment-tilting experiment (Monnier et al. 2009\nocite{2009ApJ...700..491M}),
 it displays a nearly symmetric source for baselines $<10$\,m  with a $\rm FWHM\approx120\,mas$ (Fig.\,\ref{ste}).
A spherical approximation produces the largest source sizes  at maximum silicate opacity and 
dispersed visibilities are expected to display a minimum around 9.7$\mu$m. Since this is not the case, we 
discard this scenario.

CRL\,2136 has a near-IR polarization disk (Minchin et al. 1991; Murakawa et al. 2008), and
the employed VLTI baseline is along this disk, perpendicular to the
outflow direction. Our second modelling approach is to explore the suitability of {\it dusty} geometries
which consist of a disk and an envelope with  an outflow cavity. The observed SED is compared to the grid of pre-computed SEDs,
 which were produced by 2.5D dust radiative transfer (RT) models (Robitaille et al. 2007\nocite{2007ApJS..169..328R}). Subsequently, we compute with the
same RT code monochromatic images (for code details, 
see Whitney et al. 2003\nocite{2003ApJ...591.1049W}).  This is a common
and very efficient way of performing RT modelling
(see e.g. Linz et al. 2009\nocite{2009A&A...505..655L}; Follert et al. 2010\nocite{2010A&A...522A..17F}).  
We find that grid models that fit the SED for reasonable inclinations  have a dusty disk (web-grid models \#3005338 and \#3012893); these SED fits are not presented here. The dust disk dominates the
$N$-band emission and produces pronounced mid-IR fluxes. Moreover, these disks are quite small, thus expected to have
visibilities close to unity on short baselines  (the blue dotted and red dash-dotted lines in Fig.\,\ref{ste}), something that is clearly ruled 
out by the Keck data. To roughly reproduce the Keck data and SED simultaneously, an
inner dust rim is required at about 5 to 7 times the formal dust
sublimation radius (for $T_{\rm subl}=1600$\,K). Comparing the  grid models to MIDI data, we again find that the
visibilities produce a minimum at 9.7\micron, which is not
observed.  The reason for this disagreement is the constancy of the relative contributions to the total 
flux as a function of wavelength of the disk and envelope 
(within the MIDI interferometric field of view of $\sim 0.3\arcsec$), except in the
wings of the profile where the larger envelope dominates more because
it is less affected by  extinction.

\begin{figure}
  \includegraphics[height=9cm,width=7cm,angle=90]{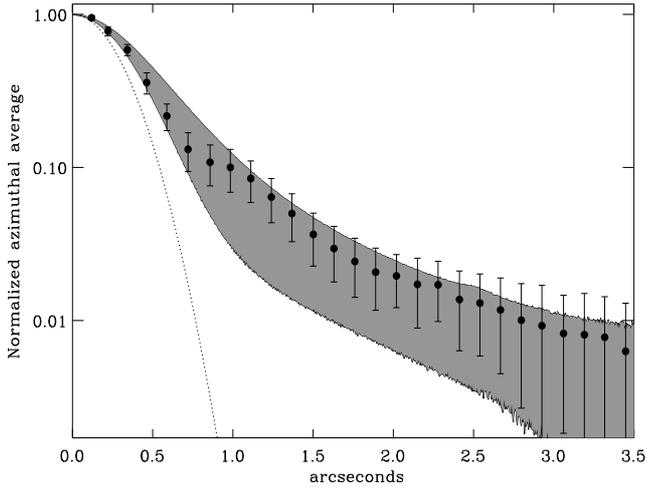}
  \caption[]{The normalized, azimuthally averaged, intensity profile at 24.5\,\micron\, of CRL\,2136 (dots represent mean values and error bars are the 
standard deviations per distance bin). The shaded region indicates the predictions of the envelope model of Sect.\,3.2. The dotted line represents the 
PSF with which the model was convolved.}
  \label{norm24}
\end{figure}

\begin{table*}
  {
    \caption[]{\label{modpar} Basic parameters for the dust envelope model presented in Sect.\,3.2. }
    \centering
    \begin{tabular}{cccccccccc}
      \hline
      \hline
      $M_{*}$       & $R_{*}$         & $T_{\rm eff}$ & $\dot{M}_{\rm infall}$ &  $i$\tablefootmark{1}    & P.A.\tablefootmark{2}            & $R_{\rm env}$  &  $R_{\rm in}$ &  $R_{\rm subl}$&$A_{\rm V}^{\rm env}$ \\  
      (\Msun)        & ($\rm \Rsun$)& (K)               &  ($\Msunyr$)                &  ($\degr$)                       &($\degr$)     & (AU)                &  (AU)             &  (AU)              &                                  \\                  
      \hline
      20                &  25.0             & 20\,000        &    $9.0\,10^{-4}$         & 70.0                                 &     $-62$     & $5.0\,10^{5}$ &     170.9        &  23                 &100.0                          \\
      \hline
      \hline
    \end{tabular}
    \tablefoot{
      \tablefoottext{1}{Inclination with respect to the plane of the sky;}
      \tablefoottext{2}{Position Angle defined as degrees East of North  with respect to the outflow axis}
    }
  }
\end{table*}

\begin{figure}
\center
  \includegraphics[height=7cm,width=7cm]{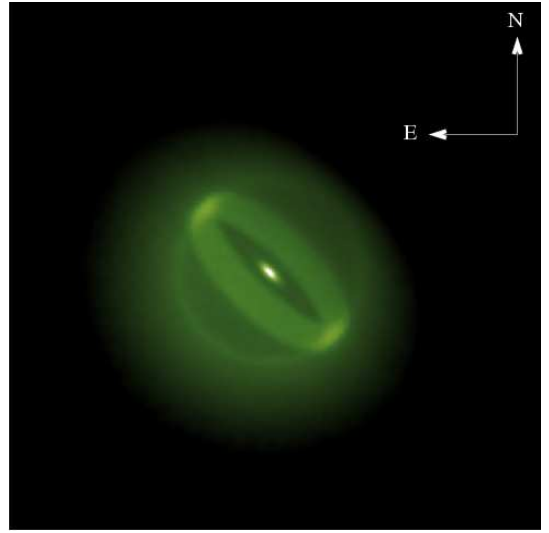}
  \caption[]{Logarithmic image at 7.5\,\micron\, of the dusty envelope with the gas disk inserted (scale is $0.5\arcsec\times0.5\arcsec$  or 1000\,AU on each side). 
Clearly visible are the dusty torus which is responsible for  a large fraction of the mid-IR radiation and the hot central parts of the disk. The faint circular emission 
is due to dust in the outflow cavities at the chosen dust rim of 170\,AU. The blue outflow cavity is to the south-east.}
    \label{dens}
\end{figure}

\subsection{Viable scenarios}
As a third approach, these considerations lead us to evaluate the significant contribution to
the total $N$-band flux, but located interior to the dust rim, of: (1) the star itself and (2) a gaseous
disk. The best examples of these two types of model are presented alongside the observations in Fig.\,\ref{spec}.  
First, we construct a dust envelope illuminated by a ZAMS star (but without disk
component). As a starting point, we use the Murakawa et
al. (2008) geometry, which has a parabolic cavity geometry  and a cavity density that is roughly a thousand
times lower than that of the envelope.  High mid-IR fluxes are
now achieved by extending the inner boundary of the envelope in the z-direction, rather than emission from the disk as previously. This gives rise to a torus-like dust
structure (Fig.\,\ref{dens}), a geometry previously suggested based on far-IR data by Harvey et al. (2000\nocite{2000ApJ...534..846H}). The inner dust rim is constrained to be
at a distance of $\sim 170$\,AU from the star, required to fit the short baselines of the Keck observations (black full and dashed lines in Fig.\,\ref{ste}). We ran a total of about 150 RT 
models and iterated to an optimal model envelope (see Table\,\ref{modpar}).  We kept the same cavity opening angle
as in the Murakawa et al. (2008) model.  The ``envelope-only'' model is also presented in Fig.\,\ref{spec} by the green dash-dotted line. We 
note that the addition  of a gaseous disk (see below) does not substantially alter the total model flux.  The  ``envelope-only'' model also reproduces the azimuthally 
averaged profile of  a spatially resolved 24.5\,\micron\,image, taken with the 8m SUBARU telescope (de Wit et al. 2009) (see Fig.\,\ref{norm24}).

With these envelope parameters at hand, we verified that, indeed,  a (hot) ZAMS object as a central illuminating source is too faint at $N$-band 
to have a significant effect on the visibilities. On the other hand,  an equally luminous, cool star {\it is} bright enough in $N$-band to 
have a significant contribution. The exchange of a hot for a cool star is motivated by work by Hosokawa \& Omukai
(2009\nocite{2009ApJ...691..823H}) and Hosokawa et al. (2010\nocite{2010ApJ...721..478H}).
These authors  illustrate that, during the main accretion phase, the central object undergoes a period of
significant swelling ($\sim 100\,\Rsun$) and has a relatively cool $T_{\rm
  eff}\sim 6000\,{\rm K}$. A cool star with the luminosity of CRL\,2136  (the red dashed line in Fig.\,\ref{spec}) has an intrinsic flux 
of a few percent of the total $N$-band flux. However, since an envelope illuminated by a cool star generates
less flux at the short end of the mid-IR  wavelength range, the fit to the SED is somewhat poorer, although 
the MIDI visibility spectrum is accurately reproduced.

Alternatively, adding to the envelope model an accretion disk interior to the
dust rim is also a viable scenario (the blue full line in Fig.\,\ref{spec}).  We then approximate the fluxes and
visibilities with an optically thick, geometrically thin, viscous disk
(Bertout et al. 1988\nocite{1988ApJ...330..350B}; Malbet et
al. 2007\nocite{2007A&A...464...43M}), and apply the line-of-sight extinction by the envelope
before the total model flux and visibilities are calculated. For a gas disk extending to the stellar surface, an
$\dot{M}_{\rm acc}$ of $3.0\,10^{-3}$\,\Msunyr\,is required. The inner edge of the disk has a temperature of
approximately 15000\,K. Clearly, the local temperature decreases if
the inner edge is at larger radii, and attains 5000\,K at 1\,AU. The
mass accretion rate needs to be a factor of two higher if the inner disk
edge is at 1\,AU.  The temperatures are consistent with the derived
distance of CO-bandhead emission in MYSOs (e.g. Wheelwright et
al. 2010\nocite{2010MNRAS.tmp.1210W}). The inferred accretion rate is
also similar to that of other MYSOs, and it is an upper limit for W33A (de Wit et al. 2010).

\section{Concluding remarks}
\label{conclusions}
We have analysed mid-IR interferometric observations of the massive young stellar object CRL 2136 obtained with the VLTI and MIDI. 
The dispersed visibilities show that there is a strong change in character
of the emitting region at $\lambda=8.5\,\mu$m.  We have found that either a cool
star or an accretion disk interior to a dusty envelope can explain the presented spatial 
and spectral observations. 
At present, neither scenario can be excluded based on the single MIDI baseline discussed in this paper, but clearly
multi baselines will help to distinguish between them. We confidently conclude that the rim
of the dust envelope is found at about 7 times the formal dust sublimation radius. The central
dust-free zone may have been evacuated by the ionized stellar wind seen  by Menten \& van der Tak
(2004\nocite{}), which occupies a similar volume.  Our present work has illustrated the potential of MIDI to detect 
compact emission in massive YSOs, which is of particular interest 
given the large numbers of massive YSOs that can be studied in this way. 
Mid-IR interferometry can therefore provide a significant contribution to the characterisation of the accretion environment of massive young stellar objects.  

\begin{acknowledgements}
It is a pleasure to thank John D. Monnier for providing the Keck data and Andrea Isella for discussions on the topic. We would 
like to express our gratitude to the anonymous referee for her/his swift responses.
\end{acknowledgements}

\small
\bibliographystyle{aa}

\end{document}